\documentclass[12pt]{ussci}
\pdfoutput=1
\usepackage{enumitem}
\setlist{noitemsep}

\usepackage{xmpincl}
\includexmp{CC_Attribution_4.0_International}

\hypersetup{
    pdftitle={Autoignition of Methyl Valerate at Low to Intermediate Temperatures and Elevated Pressures in a Rapid Compression Machine},%
    pdfauthor={Bryan W. Weber},%
}

\usepackage[version=4]{mhchem}
\usepackage{siunitx}
\usepackage{graphicx}
\usepackage{pgf}
\usepgflibrary{fpu}

\usepackage[tableposition=top]{caption}
\usepackage{booktabs}
\usepackage{mathtools}
\usepackage{microtype}

\newcommand*{\logten}{\mathop{\log_{10}}}

\usepackage{textcomp}
\DeclareUnicodeCharacter{2212}{\textendash}

\usepackage[capitalize]{cleveref}
\newcommand\papertopic{Reaction Kinetics}
\DeclareSIUnit\torr{torr}

\title{ Autoignition of Methyl Valerate at Low to Intermediate Temperatures and Elevated Pressures in a Rapid Compression Machine }

\author[1*]{Bryan W.\ Weber}
\author[1]{Justin Bunnell}
\author[2]{Kamal Kumar}
\author[1]{Chih-Jen Sung}
\affil[1]{Department of Mechanical Engineering, University of Connecticut, Storrs, CT, USA}
\affil[2]{Department of Mechanical Engineering, University of Idaho, Moscow, ID, USA}
\affil[*]{Corresponding Author: \email{bryan.weber@uconn.edu}}

\begin{document}
\maketitle

\begin{abstract} 
    Methyl valerate (\ce{C6H12O2}, methyl pentanoate) is a methyl ester and a
    relevant surrogate component for biodiesel. In this work, we present
    ignition delays of methyl valerate measured using a rapid compression
    machine at a range of engine-relevant temperature, pressure, and equivalence
    ratio conditions. The conditions we have studied include equivalence ratios
    from \numrange{0.25}{2.0}, temperatures between \SIlist{680;1050}{\K}, and
    pressures of \SIlist{15;30}{\bar}. The ignition delay data demonstrate a
    negative temperature coefficient region in the temperature range of
    \SIrange[range-phrase={--}]{720}{800}{\K} for both \(\phi=2.0\),
    \SI{15}{\bar} and \(\phi=1.0\), \SI{30}{\bar}, with two-stage ignition
    apparent over the narrower temperature ranges of
    \SIrange[range-phrase={--}]{720}{760}{\K} for the lower pressure and
    \SIrange[range-phrase={--}]{740}{760}{\K} at the higher pressure. In
    addition, the experimental ignition delay data are compared with simulations
    using an existing chemical kinetic model from the literature. The
    simulations with the literature model under-predict the data by factors
    between \numlist{2;10} over the entire range of the experimental data. To
    help determine the possible reasons for the discrepancy between simulations
    and experiments, a new chemical kinetic model is developed using the
    Reaction Mechanism Generator (RMG) software. The agreement between the
    experimental data and the RMG model is improved but still not satisfactory.
    Directions for future improvement of the methyl valerate model are
    discussed.
\end{abstract}


\section{Introduction}\label{sec:introduction}

For transportation applications, biodiesel is an important constituent in
improving environmental friendliness of fuels, due to its renewability when
produced from sustainable agricultural crops and its ability to reduce emissions
relative to conventionally fueled engines~\autocite{Hoekman2012}. A recent
review paper summarizes the work on methyl esters relevant to biodiesel
combustion~\autocite{Coniglio2013}. Autoignition of methyl butanoate (MB,
\ce{C5H10O2}) has been well-studied in both shock tube and rapid compression
machine experiments, and readers are referred to the review of
\textcite{Coniglio2013} for further details. The prevalence of MB data in the
literature is largely due to the early identification of MB as a potential
surrogate fuel for biodiesel~\autocite{Fisher2000}. However, the experiments
have shown that MB may not be an appropriate surrogate for biodiesel, due to its
lack of negative temperature coefficient (NTC) behavior, a requirement for a
suitable biodiesel surrogate~\autocite{Coniglio2013}.

Larger methyl esters such as methyl valerate (MV, \ce{C6H12O2}, methyl
pentanoate) have also been studied as possible biodiesel surrogates.
\textcite{Hadj-Ali2009} used a rapid compression machine (RCM) to study the
autoignition of several methyl esters including MV. Although MV exhibited
two-stage ignition in this study, little additional research has been done on
its oxidation. \textcite{Korobeinichev2015} studied MV in premixed laminar
flames and extended a detailed high temperature chemical kinetic model to
include MV and methyl hexanoate. \textcite{Dmitriev2015} added MV to
n-heptane/toluene fuel blends to determine the resulting intermediate species in
premixed flames using a flat burner at \SI{1}{atm} and an equivalence ratio of
1.75. The addition of MV helped reduce soot forming intermediates including
benzene, cyclopentadienyl, acetylene, propargyl, and
vinylacetylene~\autocite{Dmitriev2015}. \textcite{Hayes2009} computationally
examined the peroxy radical isomerization reactions for MV to better understand
the low temperature reaction pathways. Finally, \textcite{Dievart2013} used
diffusion flames in the counterflow configuration to determine extinction limits
for a number of methyl esters, including MV, and validated a detailed kinetic
model with the experimental data.

This work provides additional data for the autoignition of MV. Data is collected
in a RCM under engine relevant conditions spanning
from \SIrange{15}{30}{\bar}, equivalence ratios from \numrange{0.25}{2.0}, and
temperatures from \SIrange{682}{1048}{\K}. The NTC region of MV is mapped out to
provide additional information on the fidelity of using MV as a biodiesel
surrogate.

\section{Experimental Methods}\label{sec:experimental-methods}

The RCM used in this study is a single piston arrangement and is pneumatically
driven and hydraulically stopped. The device has been described in detail
previously~\autocite{Mittal2007a} and will be described here briefly for
reference. The end of compression (EOC) temperature and pressure (\(T_C\) and
\(P_C\) respectively), are independently changed by varying the overall
compression ratio, initial pressure \((P_0)\), and initial temperature \((T_0)\)
of the experiments. The primary diagnostic on the RCM is the in-cylinder
pressure. The pressure data is processed by a Python package called UConnRCMPy~\autocite{uconnrcmpy}, which calculates \(P_C\), \(T_C\), and the ignition
delay(s). The definitions of the ignition delays are shown in
\cref{fig:ign-delay-def}. The time of the EOC is defined as the maximum of the
pressure trace prior to the start of ignition and the ignition delays are
defined as the time from the EOC until local maxima in the first time derivative
of the pressure. Each experimental condition is repeated at least five times to
ensure repeatability of the data. As there is some random scatter present in the
data, the standard deviation \((\sigma)\) of the ignition delays from the runs
at a given condition is computed. In all cases, \(\sigma\) is less than
\SI{10}{\percent} of the mean value of the overall ignition delay.

In addition to the reactive experiments, non-reactive experiments are conducted
to determine the influence of machine-specific behavior on the experimental
conditions and permit the calculation of the EOC temperature via the isentropic
relations between pressure and temperature~\autocite{Lee1998}. The EOC
temperature is calculated by the procedure described in
\cref{sec:computational-methods}.

\begin{figure}[htb]
    \begin{minipage}[t]{0.48\textwidth}
        \centering
        \resizebox{\linewidth}{!}{\input{figures/ignition_delay_definition.pgf}}
        \caption{Definition of the ignition delays used in this work. The
        experiment in this figure was conducted for a \(\phi=2.0\) mixture with
        \mbox{\(\ce{Ar}/(\ce{N2}+\ce{AR})=0.5\)}, \(P_0=\SI{0.7694}{\bar}\),
        \(T_0=\SI{373}{\K}\), \(P_C=\SI{14.94}{\bar}\), \(T_C=\SI{723}{\K}\),
        \(\tau=\SI{27.44\pm0.99}{\ms}\), \(\tau_1=\SI{16.57\pm0.48}{\ms}\).}
        \label{fig:ign-delay-def}
    \end{minipage}\hfill%
    \begin{minipage}[t]{0.48\textwidth}
        \centering
        \resizebox{\linewidth}{!}{\input{figures/vapor_pressure.pgf}}
        \caption{Saturated vapor pressure of MV as a function of temperature,
        plotted using the Antoine equation, \cref{eq:antoine}, with
        \(A=6.4030\), \(B=1528.69\), and \(C=52.881\).}
        \label{fig:vapor-pressure}
    \end{minipage}
\end{figure}

The RCM is equipped with heaters to control the initial temperature of the
mixture. After filling in the components to the mixing tanks, the heaters are
switched on and the system is allowed \SI{1.5}{\hour} to come to steady state.
The mixing tanks are also equipped with magnetic stir bars so the reactants are
well mixed for the duration of the experiments.

The initial temperature is chosen such that the saturated vapor pressure
\((P_{\text{sat}})\) of the fuel at the initial temperature is at least twice
the partial pressure of the fuel in the mixing tank. The Antoine equation
\begin{equation}\label{eq:antoine}
    \logten{P_{\text{sat}}} = A - \frac{B}{T - C}
\end{equation}
is used to model the saturated vapor pressure of MV as a function of
temperature, where \(A\), \(B\), and \(C\) are substance-specific coefficients.
The coefficients in \cref{eq:antoine} are determined by least squares fitting of
the data of \textcite{Ortega2003}, \textcite{vanGenderen2002}, and
\textcite{Verevkin2008} using the \verb|curve_fit()| function of
SciPy~\autocite{Jones2001} version 0.18.1. \Cref{fig:vapor-pressure} shows that
the coefficients fit with this procedure give good agreement with the
experimental data.

The mixtures considered in this study are shown in \cref{tab:mixtures}. Mixtures
are prepared in stainless steel mixing tanks. The proportions of reactants in
the mixture are determined by specifying the absolute mass of the fuel, the
equivalence ratio \((\phi)\), and the ratio of \(\ce{Ar}:\ce{N2}\) in the
oxidizer. Since MV is a liquid with a relatively small vapor pressure at room
temperature, it is injected into the mixing tank through a septum. Proportions
of \ce{O2}, \ce{Ar}, and \ce{N2} are added manometrically at room temperature.

\begin{table}[htb]
    \centering
    \caption{Mixtures considered in this work}
    \begin{tabular}{Sccccc}
        \toprule
        {\(\phi\)} & \multicolumn{4}{c}{Mole Fraction (purity)} & \ce{Ar}/(\ce{N2} + \ce{Ar}) \\
        \cmidrule{2-5}
         & \ce{MV} (\SI{100}{\percent}) & \ce{O2} (\SI{99.994}{\percent}) & \ce{Ar} (\SI{99.999}{\percent}) & \ce{N2} (\SI{99.999}{\percent}) &  \\
        \midrule
        0.25 & 0.0065 & 0.2087 & 0.7848 & 0.0000 & 1.0 \\
        0.5 & 0.0130 & 0.2074 & 0.7798 & 0.0000 & 1.0 \\
        1.0 & 0.0256 & 0.2047 & 0.7697 & 0.0000 & 1.0 \\
        1.0 & 0.0256 & 0.2047 & 0.3849 & 0.3848 & 0.5 \\
        2.0 & 0.0499 & 0.1996 & 0.0000 & 0.7505 & 0.0 \\
        2.0 & 0.0499 & 0.1996 & 0.3752 & 0.3753 & 0.5 \\
        \bottomrule
    \end{tabular}
    \label{tab:mixtures}
\end{table}


\section{Computational Methods}\label{sec:computational-methods}
\subsection{RCM Modeling}\label{sec:experimental-modeling}

The Python 3.5 interface of Cantera~\autocite{cantera} version 2.2.1 is used for
all simulations in this work. Detailed descriptions of the use of Cantera for
these simulations can be found in the work of \textcite{Weber2016a} and
\textcite{Dames2016}; a brief overview is given here. As mentioned in
\cref{sec:experimental-methods}, non-reactive experiments are conducted to
characterize the machine-specific effects on the experimental conditions in the
RCM. This pressure trace is used to compute a volume trace by assuming that the
reactants undergo a reversible, adiabatic, constant composition (i.e.,
isentropic) compression during the compression stroke and an isentropic
expansion after the EOC. The volume trace is applied to a simulation conducted
in an \verb|IdealGasReactor| in Cantera~\autocite{cantera} using the CVODES
solver from the SUNDIALS suite~\autocite{Hindmarsh2005}. The ignition delay from
the simulations is defined in the same manner as in the experiments. The time
derivative of the pressure in the simulations is computed by second order
Lagrange polynomials, as discussed by \textcite{Chapra2010}.

To the best of our knowledge, there are three mechanisms for MV combustion
available in the literature. The first two, by \textcite{Korobeinichev2015} and
\textcite{Dmitriev2015}, were developed to simulate flames, and do not include
the low-temperature chemistry necessary to simulate the conditions in these
experiments. The third model was developed by \textcite{Dievart2013} and
includes low-temperature chemistry of MV, although it was only validated by
comparison with flame extinction limits.
The detailed \textcite{Dievart2013} model includes 1103 species and 7557
reactions.

\subsection{Reaction Mechanism Generator}\label{sec:reaction-mechanism-generator}

In addition to using a mechanism from the literature, we investigate the use of
an automatic mechanism generator, the open-source Reaction Mechanism Generator
(RMG)~\autocite{Allen2012} version 1.0.4. The Python version of RMG is used,
which requires Python 2.7, and version 1.10.0 of the RMG database is used. The
final RMG model contains 483 species and 19990 reactions. Note that the number
of species is much lower than the \textcite{Dievart2013} model because the RMG
model focuses on only one fuel (MV), but the number of reactions is
substantially higher.

\section{Experimental Results}\label{sec:experimental-results}

\begin{figure}[htb]
    \centering
    \input{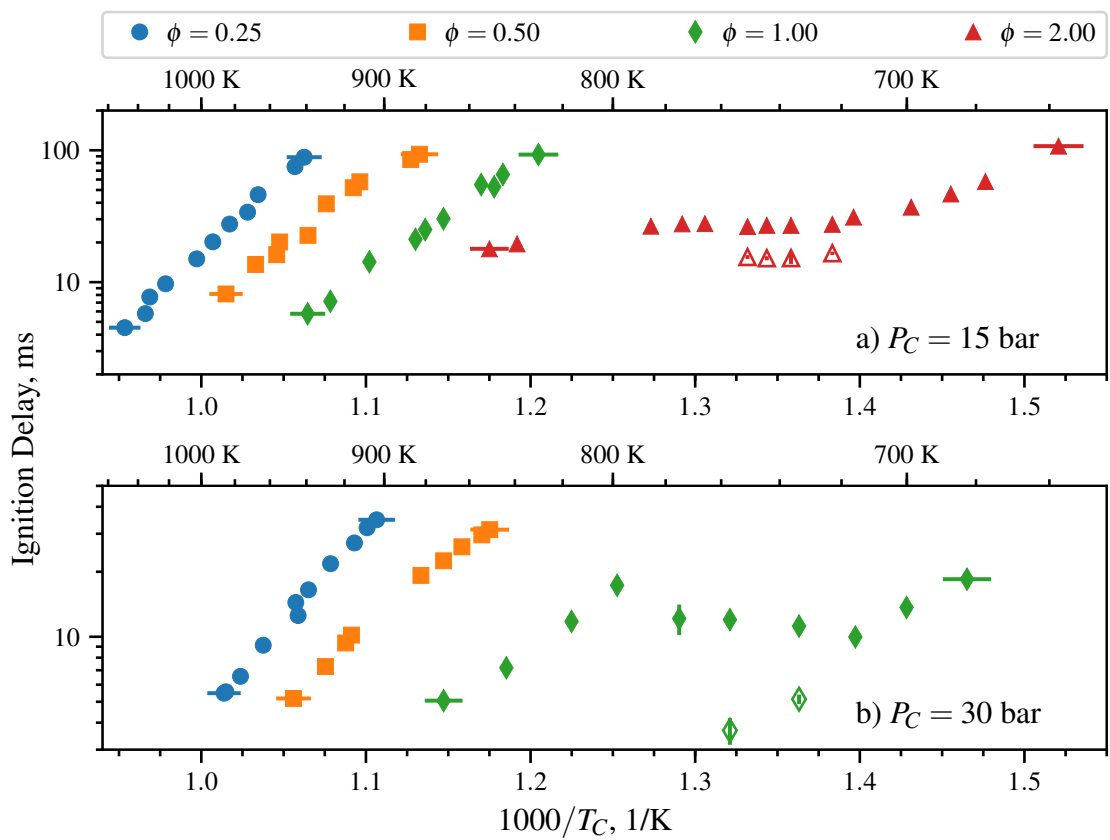}
    \caption{Ignition delays of MV as a function of inverse temperature. Filled
    points are the overall ignition delays and hollow points are the first stage
    ignition delays. a) \SI{15}{\bar}. b) \SI{30}{\bar}}
    \label{fig:ignition-delays}
\end{figure}

\Cref{fig:ignition-delays} shows the ignition delay results measured in this
study. Filled markers denote the overall ignition delay and hollow markers
indicate the first-stage ignition delay. Vertical error bars are drawn on the
symbols to represent the uncertainty in the ignition delay; for many of the
experiments, the uncertainty is approximately the same size as the data point,
so the error bar is hidden. Horizontal error bars are shown on the first and
last points of each equivalence ratio indicating the estimated uncertainty in
the EOC temperature of $\pm\SI{1}{\percent}$~\autocite{Weber2015}.
\cref{fig:ignition-delays}a shows the results for a compressed pressure of
\SI{15}{\bar}, while \cref{fig:ignition-delays}b shows the results for a
compressed pressure of \SI{30}{\bar}. Note that $\phi=2.0$ results were not
collected for \SI{30}{\bar}, so there are no red data points in
\cref{fig:ignition-delays}b.

It can be seen from \cref{fig:ignition-delays} that the ignition delays for the
\(\phi=0.25\text{ and } 0.5\) mixtures do not show an NTC region of the ignition
delay for both of the pressures studied in this work. However, the $\phi=1.0$
mixture shows an NTC region at $P_C=\SI[number-unit-product={\ }]{30}{\bar}$
between approximately \SIlist{720;800}{\K}, with measured first-stage ignition
delays at \SIlist{734;757}{\K}. In addition, the $\phi=2.0$ mixture shows an NTC
region of ignition delay at \SI{15}{\bar} from approximately
\SIrange{720}{775}{\K}, with measured first-stage ignition delays between
\SIlist{720;750}{\K}.

\Cref{fig:pressure-traces} shows the pressure traces for selected experiments at
\(\phi=1.0\), \(P_C =\SI[number-unit-product={\ }]{30}{\bar}\). The three
reactive pressure traces shown are at the low-temperature end of the NTC (blue,
\SI{700}{\K}), one case with two-stage ignition (orange, \SI{734}{\K}), and one
case near the high-temperature limit of the NTC region (green, \SI{775}{\K}).
Also shown is the non-reactive pressure trace for the \SI{700}{\K} case (red).
By comparing the \SI{700}{\K} pressure trace with the non-reactive pressure
trace, it can be seen that there is substantial heat release prior to main
ignition as measured by the deviation of the reactive pressure trace from the
non-reactive trace. However, there is only one peak in the time derivative of
the pressure, so no first-stage ignition delay is defined for this case. It can
also be seen in \cref{fig:pressure-traces} that the \SI{775}{\K} case shows some
heat release prior to ignition, although again there is only one peak in the
time derivative of the pressure. Furthermore, the heat release at \SI{775}{\K}
appears to be more gradual than at the lowest temperature.

\begin{figure}[htb]
    \begin{minipage}[t]{0.48\textwidth}
        \centering
        \resizebox{\linewidth}{!}{\input{figures/pressure-traces.pgf}}
        \caption{Selected pressure traces around the NTC region of ignition delay for $\phi=1.0$}
        \label{fig:pressure-traces}
    \end{minipage}\hfill%
    \begin{minipage}[t]{0.48\textwidth}
        \centering
        \resizebox{\linewidth}{!}{\input{figures/simulation-comparison.pgf}}
        \caption{Comparison of experimental and simulated results for $\phi=1.0$}
        \label{fig:simulation-comparison}
    \end{minipage}
\end{figure}

\section{Computational Results}\label{sec:computational-results}

\Cref{fig:simulation-comparison} compares experimentally measured overall
ignition delays with ignition delays computed with the detailed model of
\textcite{Dievart2013} for the $\phi=1.0$ experiments. Results for the other
equivalence ratios are similar to these results, so are not shown here. It is
important to note that the model of \textcite{Dievart2013} was not validated for
MV ignition delays, only for extinction strain rates. At \SI{15}{\bar}, the
model tends to under-predict the ignition delay and predicts an NTC region that
is not present in the experiments. At \SI{30}{\bar}, the model predicts the
low-temperature ignition delays well, but does not predict the NTC region found
experimentally.

To understand the underlying reasons for the disagreement between the
\textcite{Dievart2013} model and the data, we constructed an additional model
using RMG (see \cref{sec:reaction-mechanism-generator}). As can be seen in
\cref{fig:simulation-comparison}, the agreement between the RMG model and the
experimental data is similar to the \textcite{Dievart2013} model for the
\SI{30}{\bar} data. At \SI{15}{\bar}, the RMG model predicts a somewhat longer
ignition delay than the model of \textcite{Dievart2013}, but still predicts an
NTC region where none is present in the experimental data.

In general, there could be three likely sources of error in the models: missing
reaction pathways, incorrect values of the reaction rates, and incorrect values
for thermodynamic properties of the species. We have noted in
\cref{sec:reaction-mechanism-generator} that the RMG model has many more
reactions than the \textcite{Dievart2013} model and the algorithm used in RMG
considers a substantial number of the possible pathways. This reduces the
possibility of missing reaction pathways affecting the model. Further detailed
studies are required to ensure that the RMG model includes all of the relevant
reaction pathways.

The second source of error may be incorrect reaction rate parameters, either
because the rates are specified incorrectly in the model (e.g., typos) or
because the rates are not well estimated by the typical analogy based-rules. It
should be noted that errors of this type may affect the model generated by
RMG---if the rates are not estimated correctly, reactions that are important in
reality may not be included in the model. Determining the accuracy of the
reaction rates used in the RMG and \textcite{Dievart2013} models requires
further detailed studies of the models. Another related source of error could be
incorrect estimation of the pressure dependence of the reaction rates, which may
be particularly important for the isomerization reactions prevalent in
low-temperature chemistry.

\begin{figure}
    \begin{minipage}[t][][b]{0.3\textwidth}
        \centering
        \includegraphics[width=\linewidth]{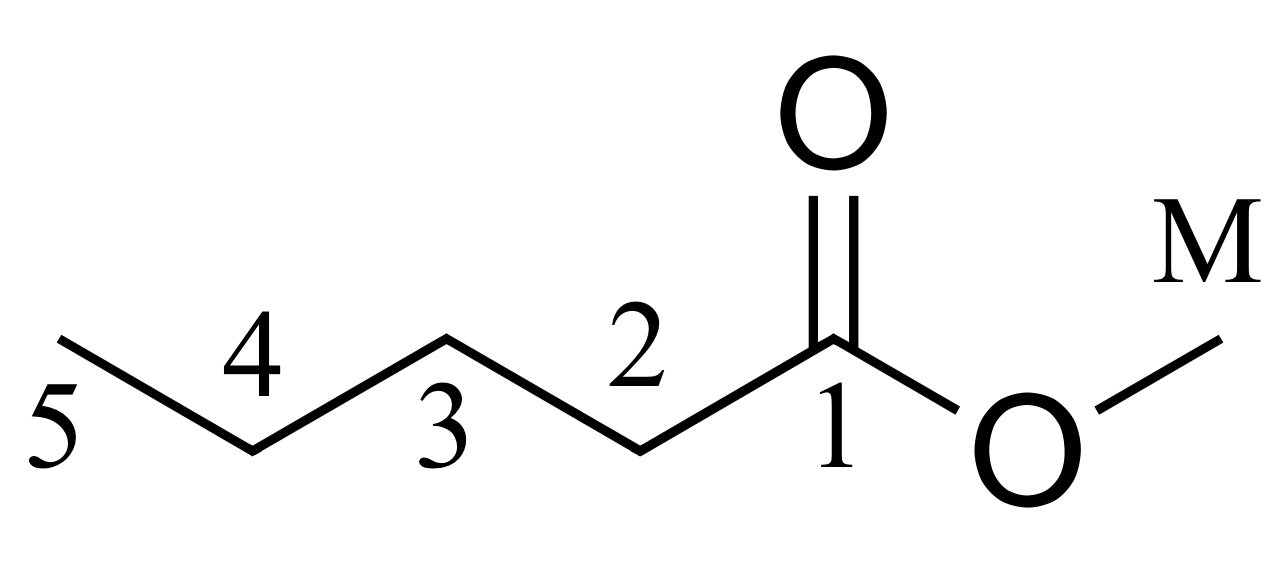}
        \caption{Structure of MV with carbon atoms labeled according to the convention used in \cref{tab:mv-radicals}}
        \label{fig:mv-structure}
    \end{minipage}\hfill%
    \begin{minipage}[t][][b]{0.68\textwidth}
        \centering
        \captionof{table}{Percent of MV destroyed to form fuel radical species with a hydrogen atom missing at the location indicated in the first column}
        \label{tab:mv-radicals}
        \begin{tabular}{cSS}
            \toprule
            Radical Site & {\textcite{Dievart2013} [\si{\percent}]} & {RMG Model [\si{\percent}]} \\
            \midrule
            2 & 29.3 & 7.4 \\
            3 & 17.5 & 36.0 \\
            4 & 17.5 & 41.1 \\
            5 & 9.4 & 3.7 \\
            M & 26.3 & 11.8 \\
            \bottomrule
        \end{tabular}
    \end{minipage}
\end{figure}

The third source of error may lie in the estimation of the thermodynamic
properties of the species, particularly their heats of formation. We have begun
to analyze the possibility of this source of error by conducting a reaction
pathway analysis to determine which radicals are formed from the breakdown of
the fuel. The following analysis is conducted for a constant volume simulation
at \SI{700}{\K}, \SI{30}{\bar}, where the rates of production of the species
have been integrated until the time of \SI{20}{\percent} fuel consumption. The
results of this analysis are shown in \cref{fig:mv-structure} and
\cref{tab:mv-radicals} for the two models. The percentages shown in the
\cref{tab:mv-radicals} are the percent of the fuel destroyed to form a
particular fuel radical by all the reactions that can form that radical.

At the relatively low temperature and high pressure condition of this analysis,
all of the fuel is destroyed by \ce{H}-atom abstractions to form the fuel
radicals shown. It can be seen that the two models have quite different
distributions of products from the first \ce{H}-abstraction reactions. The model
of \textcite{Dievart2013} predicts that \ce{H}-abstraction from the second
carbon is the most prevalent, followed closely by abstraction from the methyl
group. This is in line with the bond energies of the \ce{C-H} bonds for those
carbon atoms; we expect that the presence of the oxygen atoms will cause
hydrogen abstraction at the nearby carbons to be favored. However, the RMG model
predicts that radicals in the middle of the carbon chain will be primarily
formed. The cause of this discrepancy is under investigation, but it may be
caused by the estimation of thermodynamic properties of the radicals.

\section{Conclusions}\label{sec:conclusions}

In this study, we have measured ignition delays for methyl valerate over a wide
range of engine-relevant pressures, temperatures, and equivalence ratios. An NTC
region of the ignition delay and two-stage ignition were recorded for pressures
of \SI{15}{\bar} at \(\phi=2.0\) and \SI{30}{\bar} at \(\phi=1.0\). A detailed
chemical kinetic model available in the literature was unable to reproduce the
experimental results, so a new model was constructed using the Reaction
Mechanism Generator software. Although the new model contains many more
reactions than the literature model, it is still unable to predict the
experimental ignition delays satisfactorily. Possible reasons for the
discrepancy include missing reaction pathways, incorrect rate estimates, and
incorrect thermodynamic property estimates. Future work will include
investigation of the discrepancies between models and experiments to further
understand the autoignition kinetics of methyl valerate.

\printbibliography

\end{document}